\documentclass{aastex63}
\usepackage{epstopdf}
\usepackage{graphicx}
\received{January 19, 2021}
\revised{March 09, 2021}
\accepted{March 13, 2021}
\submitjournal{ApJL}

\shorttitle{3D magnetic topology of on-disk prominence bubble}
\shortauthors{Guo et al.}

\begin{document}

\title{Reconstructing 3D Magnetic Topology of On-disk Prominence Bubbles from Stereoscopic Observations}

\correspondingauthor{Jun Zhang}
\email{zjun@ahu.edu.cn}
\correspondingauthor{Yijun Hou}
\email{yijunhou@nao.cas.cn}

\author[0000-0001-9893-1281]{Yilin Guo}
\affiliation{CAS Key Laboratory of Solar Activity, National Astronomical Observatories,
Chinese Academy of Sciences, Beijing 100101, China}
\affiliation{School of Astronomy and Space Science, University of Chinese Academy of Sciences, Beijing 100049, China}

\author[0000-0002-9534-1638]{Yijun Hou}
\affiliation{CAS Key Laboratory of Solar Activity, National Astronomical Observatories,
Chinese Academy of Sciences, Beijing 100101, China}
\affiliation{School of Astronomy and Space Science, University of Chinese Academy of Sciences, Beijing 100049, China}

\author[0000-0001-6655-1743]{Ting Li}
\affiliation{CAS Key Laboratory of Solar Activity, National Astronomical Observatories,
Chinese Academy of Sciences, Beijing 100101, China}
\affiliation{School of Astronomy and Space Science, University of Chinese Academy of Sciences, Beijing 100049, China}

\author{Jun Zhang}
\affiliation{School of Physics and Materials Science, Anhui University, Hefei 230601, China}

\begin{abstract}

Bubbles, the semi-circular voids below quiescent prominences (filaments), have been extensively investigated in the past decade. However, hitherto the magnetic nature of bubbles cannot be verified due to the lack of on-disk photospheric magnetic field observations. Here for the first time, we find and investigate an on-disk prominence bubble around a filament barb on 2019 March 18 based on stereoscopic observations from NVST, SDO, and STEREO-A. In high-resolution NVST H$\alpha$ images, this bubble has a sharp arch-like boundary and a projected width of $\thicksim$26 Mm. Combining SDO and STEREO-A images, we further reconstruct 3D structure of the bubble boundary, whose maximum height is $\thicksim$15.6 Mm. The squashing factor Q map deduced from extrapolated 3D magnetic fields around the bubble depicts a distinct arch-shaped interface with a height of $\thicksim$11 Mm, which agrees well with the reconstructed 3D structure of the observed bubble boundary. Under the interface lies a set of magnetic loops, which is rooted on a surrounding photospheric magnetic patch. To be more persuasive, another on-disk bubble on 2019 June 10 is presented as a supplement. According to these results obtained from on-disk bubble observations, we suggest that the bubble boundary corresponds to the interface between the prominence dips (barb) and the underlying magnetic loops rooted nearby. It is thus reasonable to speculate that the bubble can form around a filament barb below which there is a photospheric magnetic patch.

\end{abstract}

\keywords{Solar atmosphere (1477); Solar filaments (1495); Solar magnetic fields (1503); Solar prominences (1519)}

\section{Introduction}

Solar prominences are cool and dense plasma structure suspended in hot and tenuous corona. In H$\alpha$ channel, they appear as bright features when observed above solar limb and appear as dark filamentary objects when observed on the solar disk, where they are referred as ``filaments'' \citep{2010SSRv..151..333M,2014LRSP...11....1P,2020RAA....20..166C}. Recent high-resolution observations reveal that dark ``bubbles'' with bright arch-like boundaries are formed below prominences, and small-scale upward ``plumes" are usually detected above the bubble boundaries \citep{2008ApJ...676L..89B,2010ApJ...716.1288B}. These plumes might be caused by Rayleigh-Taylor instability \citep{2010ApJ...716.1288B,2010SoPh..267...75R,2011ApJ...736L...1H,2012ApJ...746..120H}, coupled Kelvin-Helmholtz Rayleigh-Taylor instability \citep{2017ApJ...850...60B,2019ApJ...874...57M} or magnetic reconnection \citep{2012ApJ...761....9D,2014A&A...567A.123G,2015ApJ...814L..17S}.
The bubbles and plumes are suggested as a potential way to offer and accumulate magnetic flux and helicity for the overlying prominence system \citep{2011Natur.472..197B}, which could eventually lead to the eruption of the prominence and the generation of a coronal mass ejection \citep{2006ApJ...644..575Z}.

Bubbles have been extensively investigated based on high-resolution observations in the past decade \citep{2010ApJ...716.1288B,2011Natur.472..197B,2017ApJ...850...60B,2010A&A...514A..68S,2012ApJ...761....9D,2015ApJ...812...93G,2015ApJ...814L..17S}. \citet{2011A&A...531A..69L} found that the intensity of Fe XII line in bubbles is larger than that in prominence and lower than that in corona. Based on extreme-ultraviolet (EUV) observations, \citet{2011Natur.472..197B} calculated the temperature of the bubble plasmas in the range of (2.5-12)$\times$10$^{5}$ K, which is 25-120 times hotter than its overlying prominence. However, \citet{2015ApJ...814L..17S} and \citet{2019FrP.....7..218A} interpreted that the temperature also might come from the foreground or background of the bubble, which seriously influences the bubble emission. Moreover, \citet{2012ApJ...761....9D} believed that bubbles could be just gaps or windows in prominences, and there are no hot materials in bubbles.

The formation of bubbles is generally believed to closely connect with the emerging magnetic flux under the prominences \citep{2011Natur.472..197B,2012ApJ...761....9D,2014A&A...567A.123G,2015ApJ...814L..17S}. \citet{2011Natur.472..197B} speculated that the emergence of twisted magnetic flux below prominences causes the origin of bubbles and further proposed that through magneto-thermal convection, bubbles and plumes carry hot plasmas and magnetic flux to the overlying prominence system. \citet{2012ApJ...761....9D} reproduced the magnetic topology of the bubbles by inserting a magnetic parasitic bipolar under a linear force-free magnetic flux rope. Based on observations from THEMIS telescope, \citet{2016ApJ...826..164L} obtained that the magnetic field strength in bubbles is larger than that in prominences and argued that magnetic pressure rather than hot plasmas drives bubbles to rise.

However, previous studies about the bubbles are all based on solar limb observations, which limits further understanding of its magnetic topology and dynamical evolution. If the bubble could be found on the solar disk (i.e., on-disk bubble), combining the direct measurement of photospheric magnetic field and multi-wavelength imaging observations, we could unveil the prominence bubble from different aspects. In this Letter, an on-disk bubble is exhibited based on observations from \emph{New Vacuum Solar Telescope} \citep[NVST;][]{2014RAA....14..705L}. Combining data from \emph{Solar Dynamics Observatory} \citep[SDO;][]{2012SoPh..275....3P} and Spacecraft-A of the \emph{Solar TErrestrial RElations Observatory} \citep[STEREO-A;][]{2008SSRv..136....5K}, we obtain stereoscopic observations of the bubble and reconstruct the three-dimensional (3D) structure of its boundary. Through nonlinear force-free field (NLFFF) extrapolations, we construct the 3D magnetic topology of the bubble. In addition, another on-disk bubble is analyzed for a supplement.

\section{Observations and Data Analysis}
Checking high-resolution NVST observations in 2019, we find some suspected on-disk bubbles, two of which are studied in detail here. The first on-disk bubble was observed by NVST in H$\alpha$ channels (the line center at 6562.8 {\AA} and line wings at $\pm$0.3 {\AA}) from 05:40:57 UT to 10:00:33 UT on 2019 March 18 with a high spatial resolution of 0.{\arcsec}136 pixel$^{-1}$ and a cadence of 27 s. And the second one was observed from 05:12:50 UT to 07:42:54 UT on 2019 June 10 with a high spatial resolution of 0.{\arcsec}165 pixel$^{-1}$ and a cadence of 42 s. H$\alpha$ observations from Global Oscillation Network Group \citep[GONG;][]{1996Sci...272.1284H} are adopted as a supplement. To study the multi-wavelength appearances of the on-disk bubbles, we also apply the 171 {\AA}, 193 {\AA}, and 211 {\AA} images from Atmosphere Imaging Assembly \citep[AIA;][]{2012SoPh..275...17L} on board the SDO with a pixel size of 0.{\arcsec}6 and a cadence of 12 s. Moreover, we use the photospheric vector magnetograms with a cadence of 720 s and line-of-sight (LOS) magnetograms with a cadence of 45 s from SDO/Helioseismic and Magnetic Imager \citep[HMI;][]{2012SoPh..275..207S}. The SDO/AIA and NVST observations are carefully co-aligned by matching specific features observed simultaneously in both the AIA 171 {\AA} and NVST H-alpha channels.

In order to reconstruct the 3D structure of the bubble boundary, we also employ the STEREO-A 171 {\AA} and 195 {\AA} observations with spatial resolutions of 1.{\arcsec}587 pixel$^{-1}$ and cadences of 60 and 2.5 minutes, respectively. Then the stereoscopic observations from STEREO-A and SDO are analyzed by using a routine called ``\verb#SCC_MEASURE.pro#'' (developed by W. Thompson). After selecting the same features in both images from STEREO-A and SDO, the routine can determine the 3D coordinate of the tiepoints by the triangulation method.

For obtaining 3D magnetic topology of the on-disk bubbles, we use the ``weighted optimization'' method to perform NLFFF extrapolations \citep{2000ApJ...540.1150W,2004SoPh..219...87W,2012SoPh..281...37W} based on the HMI photospheric vector magnetic fields at 08:00 UT on 2019 March 18 and at 06:00 UT on 2019 June 10. The NLFFF extrapolations are performed in a box of 600$\times$400$\times$512 and 500$\times$456$\times$256 uniformly spaced grid points with $\bigtriangleup$x = $\bigtriangleup$y = $\bigtriangleup$z = 1{\arcsec}, respectively. Moreover, based on the extrapolation results, we calculate the squashing factor Q with method proposed by \citeauthor{2016ApJ...818..148L} (\citeyear{2016ApJ...818..148L}).

\begin{figure}[ht!]
\centering
\includegraphics [width=0.85\textwidth]{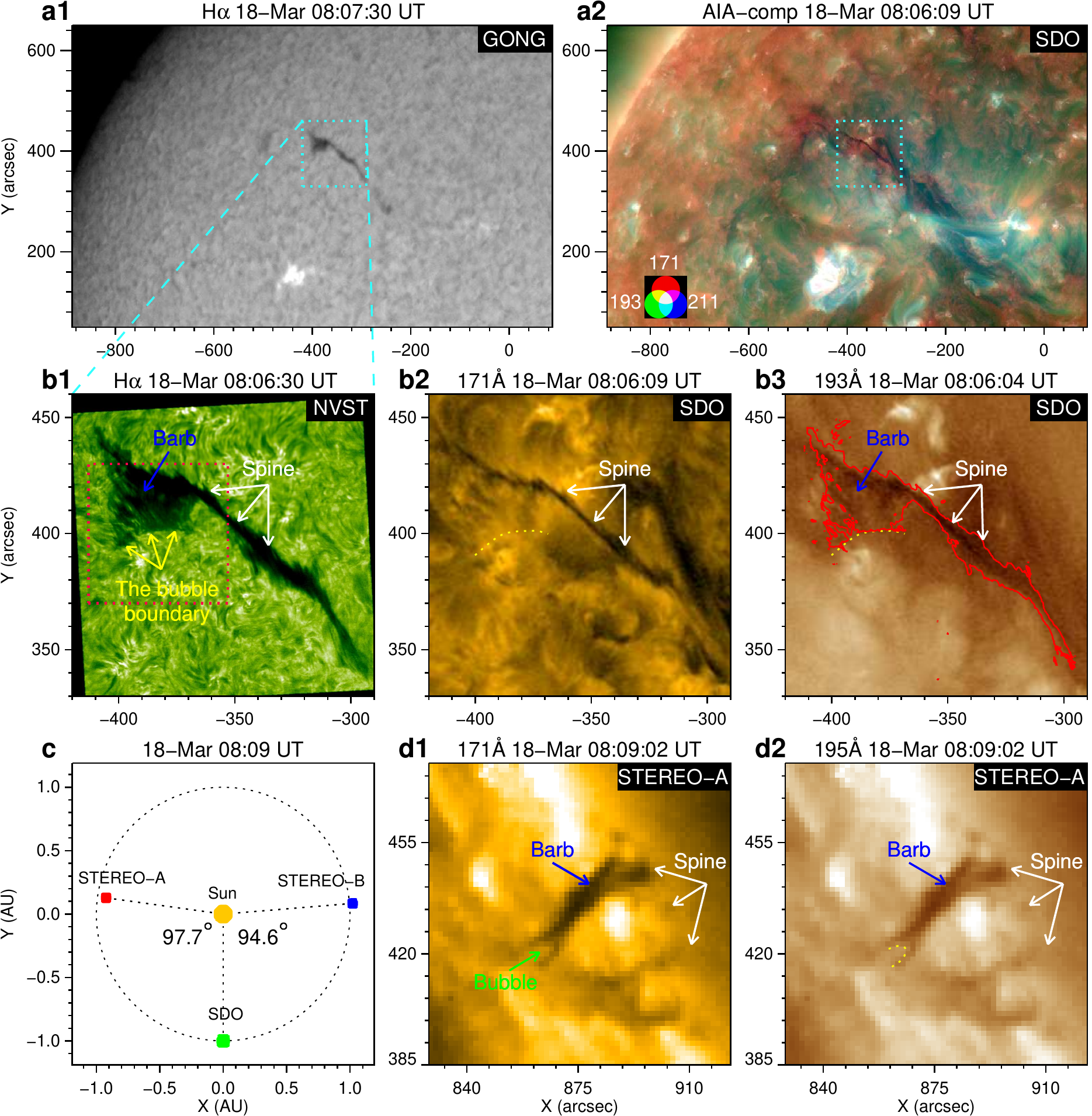}
\caption{Overview of the on-disk bubble on 2019 March 18. (a1)-(a2): GONG H$\alpha$ image and SDO/AIA composite image of 171 {\AA}, 193 {\AA}, and 211 {\AA} channels displaying the host filament (prominence) of the bubble. The cyan squares in panels (a1)-(a2) outline the FOV of panels (b1)-(b3). (b1)-(b3): NVST H$\alpha$, SDO 171 {\AA}, and SDO 193 {\AA} images exhibiting the bubble. The yellow, blue and white arrows point to the bubble boundary, the barb and the filament spine, respectively. The red square shows the FOV of Figure 2. The yellow dotted curves is the duplication of the bubble boundary in panel (b1). The red contour outlines the filament in panel (b1). (c): Relative locations of STEREO-A and SDO at 08:06 UT. (d1)-(d2): STEREO-A 171 {\AA} and 195 {\AA} images exhibiting the bubble from another view. The yellow curve outlines the bubble boundary. An animation of NVST H$\alpha$, SDO 193 {\AA}, and STEREO-A 175 {\AA} images, covering 05:40:57 UT to 10:00:33 UT on 2019 March 18, is available online.
}
\label{fig1}
\end{figure}

\begin{figure}[ht!]
\centering
\includegraphics [width=0.85\textwidth]{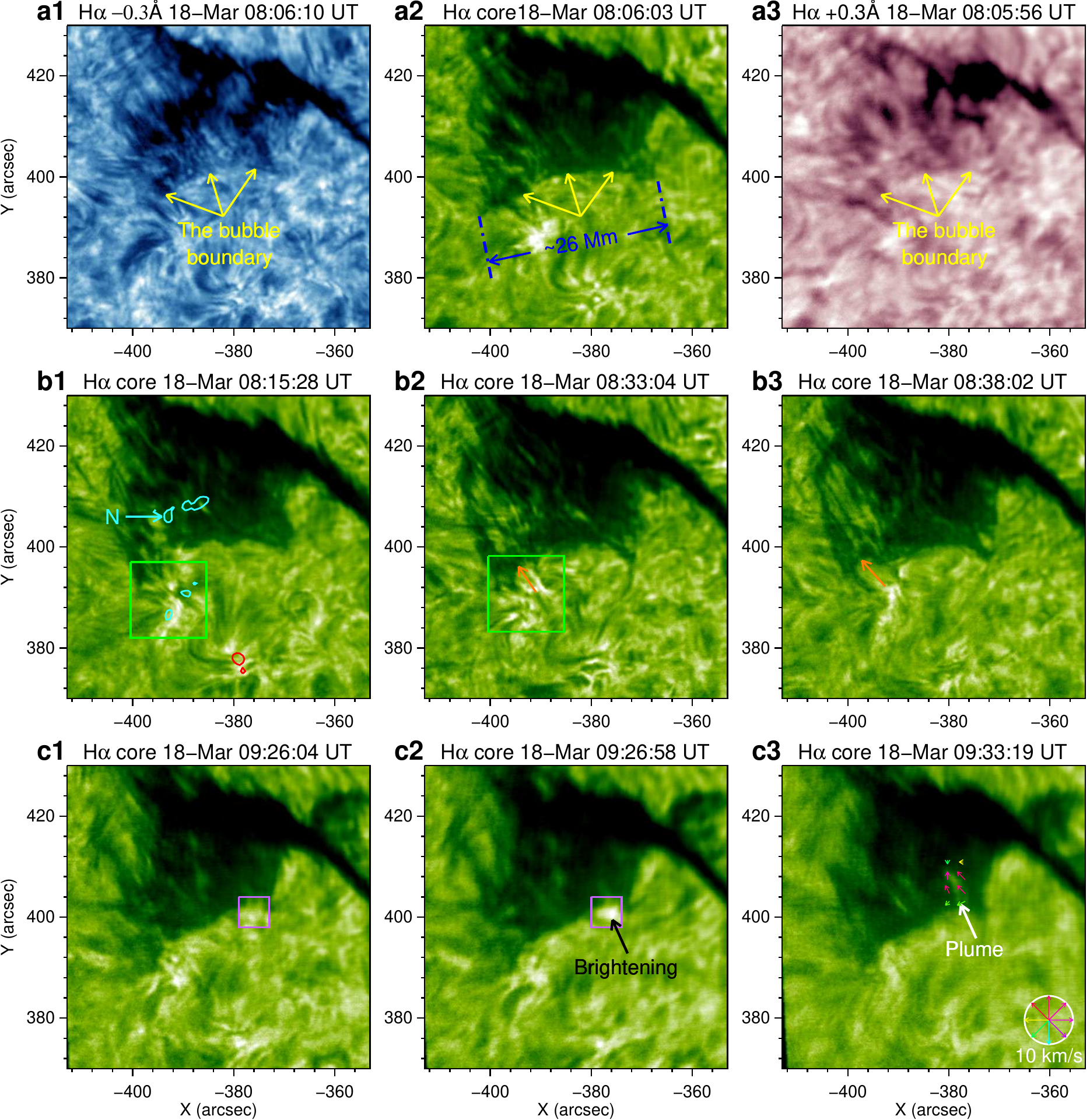}
\caption{Evolution process of the on-disk bubble. (a1)-(a3): NVST H$\alpha$ -0.3 {\AA}, line core, and +0.3 {\AA} images presenting the bubble with a sharp arch-like boundary (the yellow arrows). (b1)-(b3): H$\alpha$ line core images displaying the bubble dynamic tied to a surrounding bright structure. The cyan and red contours are the -100 and 100 G levels of the corresponding magnetic field, respectively. The orange arrows show the boundary disturbed by the bright structure (marked by green squares). (c1)-(c3): A plume (see the white arrow) driven by a transient brightening near the boundary. The pink squares highlight the region before and after brightening. The colored arrows in (c3) represent the horizontal velocity fields of the plume materials derived from H$\alpha$ images by LCT method.
}
\label{fig2}
\end{figure}

\begin{figure}[ht!]
\centering
\includegraphics [width=0.85\textwidth]{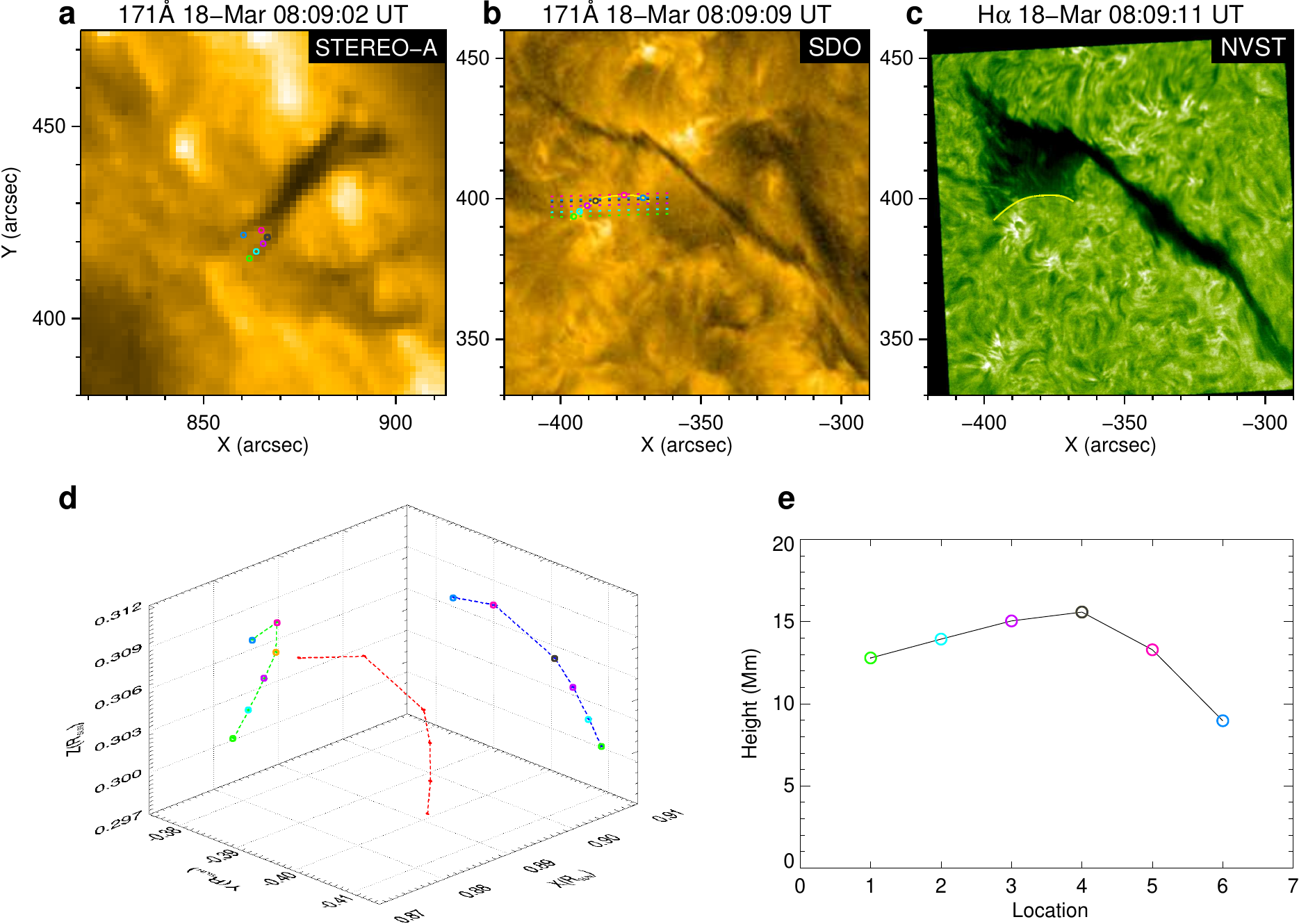}
\caption{3D structure of the on-disk bubble boundary. (a)-(c): STEREO-A 171 {\AA}, SDO 171 {\AA}, and NVST H$\alpha$ images displaying the bubble. The colored circles indicate the locations of six features selected along the bubble boundary. The colored horizontal lines in (b) are ``epipolar lines''. The yellow curve in (c) outlines the bubble boundary and then is duplicated to (b). (d): Reconstructed 3D structure of the bubble boundary (see the red curve). The green and blue curves display its projections similar to the structure observed from STEREO-A and SDO in (a) and (b), respectively. (e): Heights of the six features.
}
\label{fig3}
\end{figure}

\begin{figure}[ht!]
\centering
\includegraphics [width=0.85\textwidth]{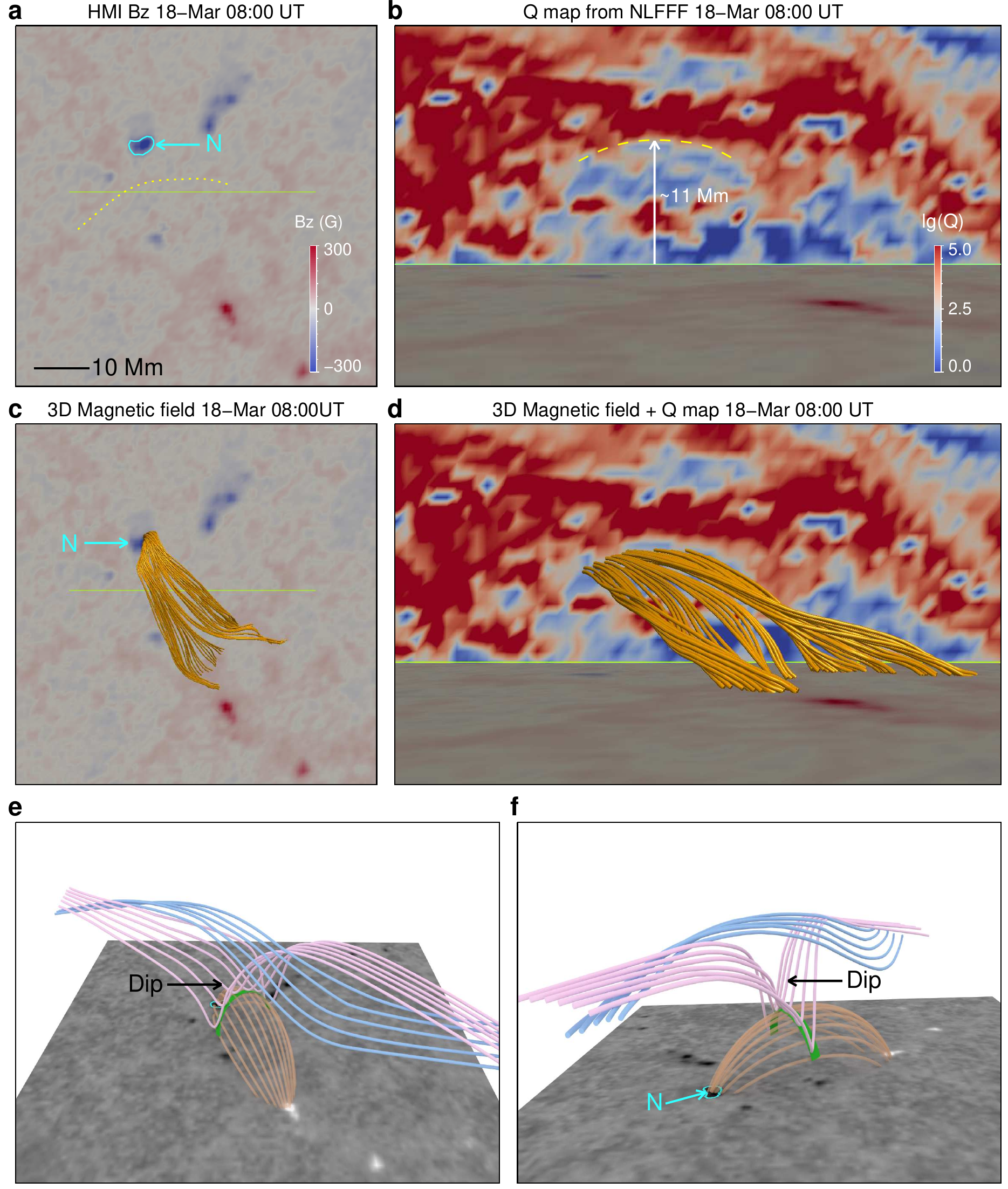}
\caption{3D magnetic topology and cartoon model of the bubble-prominence dip system. (a): The photospheric vertical magnetograms under the bubble. The yellow dotted curve outlines the approximate projected location of the bubble boundary. (b): Logarithmic Q distribution in the vertical plane above the green line denoted in (a). The yellow dashed curve traces the interface between high-Q region and low-Q region. (c): Top view of the magnetic loops revealed by NLFFF extrapolation. (d): Side view of magnetic loops combining logarithmic Q distribution. (e)-(f): The cartoon model showing the magnetic topology of the bubble-prominence dip system from different observation views. The orange and green curves represent the magnetic loops and interface, respectively. The cyan curves in (a), (e), and (f) are contours of the negative polarity (N) at +100 G.
}
\label{fig4}
\end{figure}

\begin{figure}[ht!]
\centering
\includegraphics [width=0.85\textwidth]{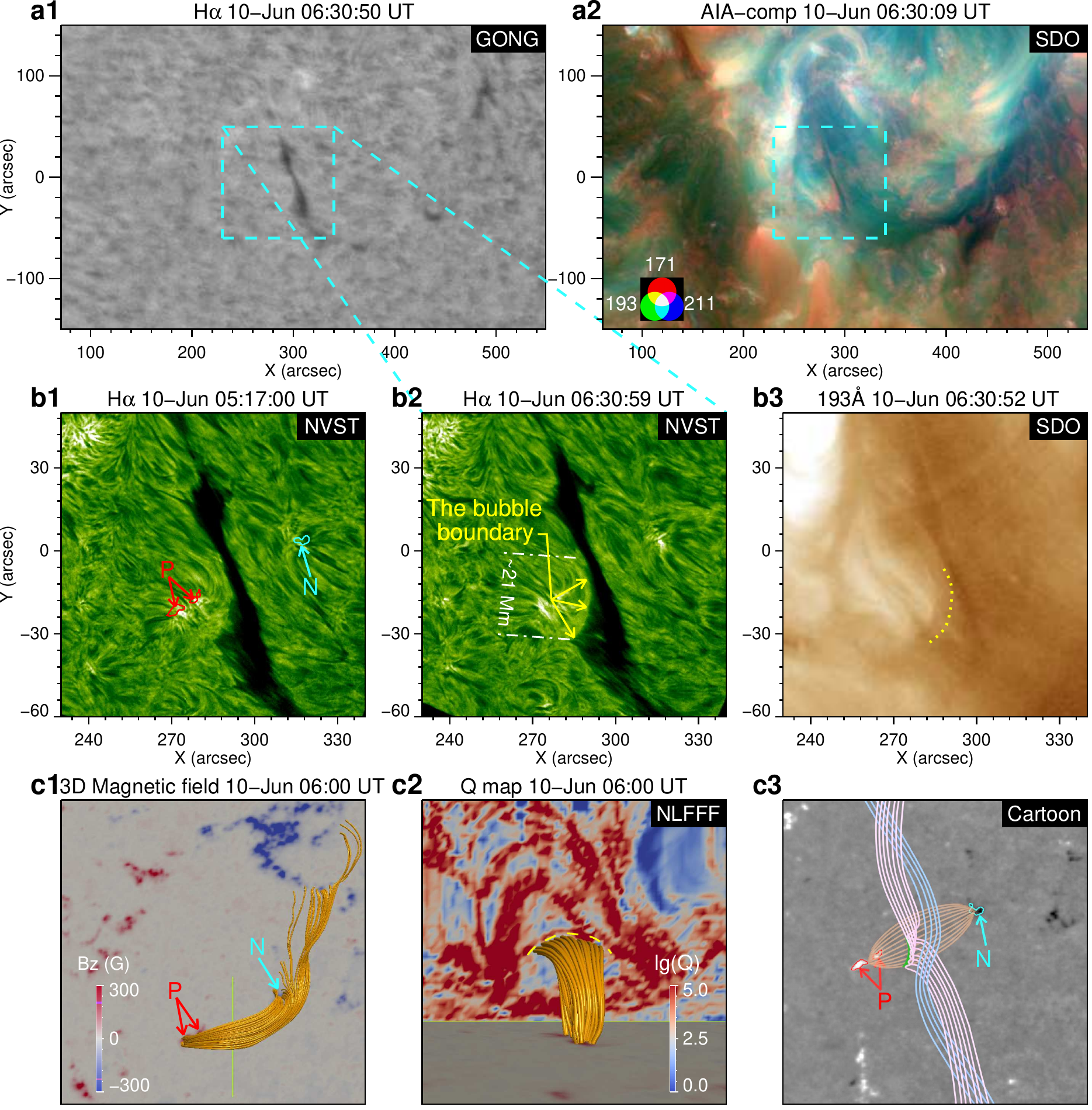}
\caption{Multi-wavelength observations, 3D magnetic topology and cartoon model of another on-disk bubble in the prominence on 2019 June 10. The features marked here are similar to those in Figures 1 and 4. An animation of NVST H$\alpha$ and SDO 193 {\AA} images, covering 05:12:50 UT to 07:42:54 UT on 2019 June 10, is available online.
}
\label{fig5}
\end{figure}

\section{Results}

The first suspected on-disk bubble was observed at (N18{\degr}, E23{\degr}) around a filament barb on 2019 March 18 (Movie 1). Figures 1(a1) and (a2) show the host filament (prominence) of this suspected bubble. As presented in high-resolution NVST H$\alpha$ image with a smaller field of view (FOV), this suspected bubble has a sharp arch-like boundary below the barb (Figure 1(b1)). However, the barb and arch-shaped boundary are not clearly detected in AIA 171 {\AA} and 193 {\AA} images (Figures 1(b2) and (b3)). This could result from that compared with the filament spine, the barb overlying the arch-shaped boundary has smaller plasma density and lower altitude, which would cause a weaker ``emissivity blocking'' for the coronal plasma with temperatures corresponding to the 195 {\AA} and 171 {\AA} channels \citep{1998Natur.396..440Z,2003SoPh..216..159H}. To verify that this structure below a barb is indeed an on-disk prominence bubble, simultaneous observations from another view is indispensable. Fortunately, the angle between SDO and STEREO-A was 97.7{\degr} at this moment (Figure 1(c)). In STEREO-A 171 {\AA} and 195 {\AA} images, this structure is located near solar west limb and presents typical features of the prominence bubble below the barb, which appears as a vertical ``tornado" structure \citep[e.g.,][]{2012ApJ...756L..41S,2013ApJ...774..123W}, (Figures 1(d1) and (d2)). Note that it could be the line-of-sight integration effect \citep{2011ApJ...739...43L,2018ApJ...867..115G} that causes the phenomenon that the bubble boundary is visible in the STEREO-A EUV observations but invisible in the SDO EUV observations.

As exhibited in Figure 2, during 05:40:57 UT to 10:00:33 UT, the on-disk bubble performed dynamical behaviors in NVST H$\alpha$ observations (marked by the red square in Movie 1). The sharp arch-like boundary of the bubble repeatedly appeared around 06:54 UT, 08:06 UT, 08:39 UT, and 09:04 UT, respectively. For each time, the sharp boundary could last about ten minutes and then was followed by processes of perturbation and collapse. Figures 2(a1)-(a3) show the bubble with a sharp arch-like boundary at around 08:06 UT in three Ha channels (i.e., two wings at $\pm$0.3 {\AA} and line core). The projected width of the bubble is $\thicksim$26 Mm. Around the bubble boundary, two types of bright structures are observed, one of which is rooted on a positive magnetic patch and intermittently disturbs the east part of the boundary (Figures 2(b1)-(b3)). Another one is the transient brightening (squares in Figures 2(c1) and (c2)) near the west part and followed by a plume (Figures 2(c3)), indicating the occurrence of magnetic reconnection around the bubble boundary \citep{2012ApJ...761....9D,2015ApJ...814L..17S}.

As mentioned above, this on-disk bubble can be observed simultaneously from Earth (SDO and NVST) view and STEREO-A view. Therefore, based on these stereoscopic observations, we reconstruct 3D structure of the bubble boundary through the routine ``\verb#SCC_MEASURE.pro#''. This routine uses an approximate ``epipolar constraint'' in locating the feature from two images observed from different views \citep{2006astro.ph.12649I}. As shown in Figure 3(a), we select a feature (colored circles) from the bubble boundary in the STEREO-A 171 {\AA} image. Then an ``epipolar line'' (the same colored lines in Figure 3(b)) would be produced passing the same feature in the SDO 171 {\AA} image. However, the emission feature of the bubble boundary in SDO 171 {\AA} image is not clear as that in NVST H$\alpha$ image. Therefore, we duplicate the clear boundary in NVST H$\alpha$ image to SDO 171 {\AA} image by a yellow curve. Then the location of this feature in SDO 171 {\AA} image is determined as the intersection between the colored line and the yellow curve in Figure 3(b). Similarly, along the bubble boundary, we select six features and finally obtain their 3D coordinates from the solar center. Converted to rectangular Cartesian coordinate system, the 3D structure of the bubble boundary is exhibited by the red curve in Figure 3(d). The Earth view is along the negative X-axis, following which we can obtain the projection of the bubble boundary (blue curve) on the Y-Z plane with similar structure as that in Figure 3(b). Likewise, along the positive Y-axis which is close to STEREO-A view, we can obtain its another projection (green curve) on the X-Z plane with similar structure as that in Figure 3(a). Heights of the six features are presented in Figure 3(e), the maximum of which is $\thicksim$15.6 Mm.

After obtaining 3D structure of the bubble boundary, we wish to further investigate the magnetic topology of the bubble. Therefore, based on the photospheric vector field of 08:00 UT on 2019 March 18, we extrapolate 3D magnetic fields around the bubble by NLFFF modeling and calculate the squashing factor Q in the vertical plane across the green line in Figure 4(a). As shown in Figure 4(b), there is an arch-shaped interface with a height of ~11 Mm outlining the boundary between the bubble and the host prominence. In the bubble semi-circular void, we can reconstruct a set of magnetic loops rooted on a surrounding photospheric magnetic patch (N) (Figures 4(c) and (d)). The connectivity of these magnetic field lines is relatively consistent, and thus a low-Q region forms under the bubble boundary. According to observations and extrapolation results, we draw a cartoon model to illustrate the magnetic topology of the bubble-prominence dip system in the bottom of Figure 4. The prominence dips pile up on the underlying magnetic field and form an arch-shaped interface, which corresponds to the bubble boundary.

As a supplement, another on-disk bubble located at (S00{\degr}, W15{\degr}) on 2019 June 10 is exhibited. Figures 5(a1) and (a2) show the host filament (prominence) of the bubble. In NVST H$\alpha$ observations, the bubble with a projection width of $\thicksim$21 Mm is presented, and it also has a sharp arch-like boundary (Figure 5(b2)). The boundary is duplicated to SDO/AIA 193 {\AA} image (Figure 5(b3)). In addition, the NLFFF extrapolation results and Q map based on the photospheric vector field of 06:00 UT on 2019 June 10 reveal that there is a interface outlining the bubble boundary, under which is a set of magnetic loops (Figures 5(c1) and (c2)), just similar to the first one. The loops anchor in a positive magnetic patch, where a bright structure is observed (Figures 5(b1)-(b2)). However, as shown in Movie 2, different from the first bubble, this bubble remained almost stable during 05:12:50 UT to 07:42:54 UT, except for several slight perturbations, and there is no direct evidence that the bubble dynamic is related to the bright structure. A cartoon model is also proposed to illustrate the magnetic topology of the bubble-prominence dip system (Figure 5(c3)).

\section{Summary and Discussion}

In this paper, we investigate an on-disk prominence bubble on 2019 March 18. Through high-resolution NVST H$\alpha$ observations, we study the radiation and evolution characteristics of the on-disk bubble from 05:40:57 UT to 10:00:33 UT. The bubble had a sharp arch-like boundary at around 06:54 UT, 08:06 UT, 08:39 UT, and 09:04 UT. Combining SDO and STEREO-A images, we reconstruct the 3D structure of the bubble boundary and obtain its maximum height of $\thicksim$15.6 Mm. The NLFFF extrapolation results reveal that the interface between prominence dips and underlying magnetic loops corresponds to the boundary of the on-disk bubble. To be more persuasive, the radiation characteristics and magnetic topology of another on-disk bubble on 2019 June 10 are presented as a supplement.

For the event on 2019 March 18, the on-disk bubble was clearly detected near the solar limb in the STEREO-A observations. According to observations from NVST, SDO, and STEREO-A, we propose that the on-disk bubble studied here is the same physical phenomenon as the prominence bubble reported before \citep{2011Natur.472..197B,2015ApJ...814L..17S}. We come to this conclusion based on the following four points: (1) the on-disk bubble observed at NVST H$\alpha$ line has a sharp arch-like boundary, just like the notable bright continuous arch-like boundary of bubbles observed from solar limb; (2) the on-disk bubble has a typical projected width of $\thicksim$26 Mm, which is consistent with bubble scales of recent observations \citep[9-50 Mm;][]{2008ApJ...687L.123D,2010ApJ...716.1288B,2017ApJ...850...60B,2019FrP.....7..218A}; (3) the on-disk bubble boundary, which undergoes repetitive formation, perturbation, and collapse within about four hours, has similar dynamic characteristics with bubbles, and plume-like structures are observed to generate at the boundary; (4) it is very intuitive that the on-disk bubble presents a bubble-like radiation characteristic when observed near the west solar limb in STEREO-A view. Another similar on-disk bubble on 2019 June 10 is also exhibited in Figure 5, which meets the first three points mentioned above. However, because it could not be simultaneously observed by the STEREO-A, we cannot further verify whether it is a real prominence bubble according to the forth point.

Previous studies about the prominence bubbles are all based on the solar limb observations or numerical simulations. And it is widely accepted that bubbles are caused by the emerging magnetic flux under prominences \citep{2011Natur.472..197B,2012ApJ...761....9D,2015ApJ...814L..17S}. However, hitherto the magnetic nature of bubbles cannot be verified due to the lack of on-disk photospheric magnetic field observations. In this work, after confirming the on-disk structures observed here are indeed bubbles, we further study their magnetic topology and formation mechanism based on observations from photospheric vector magnetic fields by employing NLFFF method. Combining NVST and SDO observations, we find that on-disk bubbles are usually detected around the filament barbs, below which there is a photospheric magnetic patch. The NLFFF extrapolation results reveal that a set of magnetic loops is rooted on this magnetic patch (Figure 4(c)). And between the set of field lines and the overlying magnetic system, there is a distinct arch-shaped interface with a height of $\thicksim$11 Mm, which is in good agreement with the 3D structure of the on-disk bubble boundary (Figure 3(d)). The height difference between the interface and the boundary might result from that the vertical plane including the interface is not completely passing through the boundary.

Combining the observations and extrapolation results, we propose a possible cartoon model of the bubble-prominence dip system, which is similar to the magnetic field models constructed by \citet{2012ApJ...761....9D}. According to NVST H$\alpha$ observations, the plasmas pile up in the prominence magnetic dips and form a barb extending from the prominence spine \textbf{\citep{1998A&A...329.1125A,2005ApJ...626..574C,2008SoPh..248...29D,2010ApJ...714..343G}}. As shown in Figures 4(e) and (f), the magnetic dips interact with the underlying magnetic loops at some locations, where an arch-shaped interface forms. And the piling-up plasmas near the interface form the bubble boundary where the barb threads terminate in NVST H$\alpha$ image. The bubble boundary could stand astride rather than parallel to \citep{2015ApJ...814L..17S} the magnetic loops, which indicates that the location of the boundary is related to both the overlying and underlying magnetic systems. From different observation views, such as SDO and STEREO-A, the boundary appears as different shapes \citep{2018ApJ...867..115G}.

Our work suggests that the bubble is just an apparent radiation structure along LOS, which might be formed around a filament (prominence) barb below which there is a photospheric magnetic patch. Moreover, similarly with the views of \textbf{\citet{2017ApJ...850...60B}}, we propose that the on-disk bubble is probably not a rare structure in on-disk observations and is important for further investigating the prominence bubble. However, in this work, the key question that whether the bubbles form from flux emergence below a pre-existing prominence is still unsettled due to the lack of sufficient H$\alpha$ and photospheric magnetic field observations of the bubble formation process. Therefore, in our upcoming study, we will explore more on-disk bubble events to solve this question. In addition, based on stereoscopic observations, we hope to obtain new insight into the dynamics of the bubble boundary layer as well.

\acknowledgments

The authors are indebted to two anonymous referees for constructive suggestions. The observations used in this paper were obtained from the NVST, SDO, STEREO-A, and GONG. This work is supported by the National Key R$\&$D Program of China (2019YFA0405000), the National Natural Science Foundation of China (12073001, 11790304, 11903050, 11773039, 11873059, 12073042 and 11790300), the Strategic Priority Research Program of the Chinese Academy of Sciences (XDB41000000), the NAOC Nebula Talents Program, the Youth Innovation Promotion Association of CAS (2017078), and Key Programs of the Chinese Academy of Sciences (QYZDJ-SSW-SLH050).



\bibliography{bubbleref}{}
\bibliographystyle{aasjournal}

\end{document}